\documentclass[lettersize,journal]{IEEEtran}
\usepackage{amsmath,amsfonts}
\usepackage{array}
\usepackage[caption=false,font=normalsize,labelfont=sf,textfont=sf]{subfig}
\usepackage{textcomp}
\usepackage{stfloats}
\usepackage{url}
\usepackage{verbatim}
\usepackage{graphicx}
\usepackage{cite}
\usepackage{multirow}
\usepackage{url}
\usepackage{footnote}
\usepackage{hyperref}

\usepackage{tabularray}

\usepackage{algorithm}
\usepackage{algpseudocode}

\usepackage{array}
\usepackage{booktabs}

\usepackage{color}
\usepackage{xcolor}

\begin{document}
 
\title{Enhancing Smart Grids with Internet of Energy: Deep Reinforcement Learning and Convolutional Neural Network}

\author{Ali Mohammadi Ruzbahani\\
University of Calgary\\
\textit{Email: ali.mohammadiruzbaha@ucalgary.ca}
}



\maketitle

\begin{abstract}
The increasing demand for electricity, coupled with the rise in greenhouse gas emissions, necessitates the integration of Renewable Energy Sources (RESs) into power grids. However, the fluctuating nature of RESs introduces new challenges in energy management. The Internet of Energy (IoE) framework provides a solution by enabling real-time monitoring, dynamic scheduling, and enhanced energy routing. This paper proposes a comprehensive approach to optimizing energy management in smart grids using Deep Reinforcement Learning (DRL) and Convolutional Neural Networks (CNN). The research focuses on three main objectives: optimizing operation scheduling, improving energy routing, and enhancing cyber-physical security. A DRL-based scheduling algorithm is developed to manage energy components effectively, while an optimized energy routing algorithm ensures efficient electricity flow. Additionally, a security framework utilizing Long Short-Term Memory (LSTM) and CNN is proposed to detect False Data Injection (FDI) attacks and electricity theft. The proposed methods aim to improve energy efficiency, reduce costs, and ensure the security of IoE-enabled power systems. This research bridges existing gaps by addressing the dynamic and complex nature of modern energy networks. The integration of these advanced technologies promises significant advancements in the reliability and efficiency of smart grids. Ultimately, this work contributes to the development of a sustainable and secure energy future.
\end{abstract}


\section{Introduction}
Electricity demand is steadily rising since industrial and economic development heavily relies on electrical energy \cite{k1,h1}. During the past ten years, electricity demand is increased uninterruptedly by an average of 3.1\% annually, which resulted in growing greenhouse gas (GHG) emissions and increases the need for new energy resources  \cite{1,h2}. According to the Energy Information Administration (EIA), 62\% of electricity production in the United States supplied by fossil energies, and 20\% average global growth has happened for these resources. Since fossil fuels are expensive and pollute, Renewable Energy Sources (RESs) are utilized dramatically  \cite{2,h3,h4}.
High penetration of RESs brings up new challenges due to supply fluctuation, uncertainties imposed by the nature of RESs and decentralized topology that originates from the wide geographical distribution of the energy resources.  \cite{3,h5}. Also, real-time monitoring of energy flow and dealing with big data generated by information infrastructures are vital to enhance the network's energy efficiency, reliability, and stability \cite{6,h7}. Furthermore, a comprehensive scheduling program is required to increase customers' monetary profit, making RESs more affordable. Finally, security is also a concern that needs to be handled since the growth of advanced information and communication technologies makes the network more vulnerable to malicious physical and cyber-attacks  \cite{k2,4}.  In a nutshell, the newly emerged challenges need to be dealt with through novel techniques that are precise and high-speed besides self-improvement capability for the upcoming modifications.
Conventional Energy Management Systems (EMSs) are not practical henceforward due to transformation in the network's topology and load curve, caused by penetration of new resources, besides existing massive data generated by smart tools. As a complex cyber-physical system, the Internet of Energy (IoE) can provide an energy management framework that facilitates the accommodation and coordination of RESs, providing real-time monitoring via embedded bidirectional communication systems. IoE is a cloud-based technology that integrates the power system with embedded metering, communication, sensing, and networking capabilities  \cite{h6}, delivering a bidirectional information structure. This is vital in an Energy Management System (EMS) to distribute electricity safely and efficiently from RESs to the system or end-users  \cite{k3,h7}.  
Operational Scheduling (OS) is one of the main objectives of the EMSs, which can be fulfilled through IoE. Scheduling the operation, maintenance, and transmission planning plays a leading role in efficient EMS. Optimized OS transforms consumers into prosumers who can participate in the energy market by selling their surplus electricity while maximizing the users' economic profit. It also accelerates the utilization of RESs, which leads to numerous technical and economic advantages \cite{7,h8}.  
In addition to facilitating OS, IoE also offers the opportunity for energy routing optimization to improve stability and reliability and to increase profit in the power network. Due to the RESs, multiple energy hubs are connected via energy routers to build an energy network, which results in the genesis of Virtual Power Plants (VPPs), one of the central concepts for improving energy efficiency [8]. Energy router is a fundamental component in EMS that regulates the direction and amount of electricity and optimizes the energy flow among all devices [9]. Therefore, optimizing energy routing in IoE needs to be considered as a crucial challenge in energy management.
Aside from numerous advantages of employing IoE in EMS, this technology is vulnerable to malicious attacks  \cite{9,k8,10}. The distributed pattern of IoE, which allows users to interact and exchange information and energy without central control, leads to many security and privacy challenges  \cite{11,e1}. Security concerns are not limited to cyber layers since the system may be physically manipulated/damaged for electricity theft or sabotages.
In summary, improving energy efficiency in the power network via the IoE concept is facing three main challenges that should be addressed: optimal OS, optimal energy routing in the different layers, and cyber/physical security concerns. Consequently, three main goals are defined for this research proposal. First, an operation scheduling algorithm is optimized using Deep Reinforcement Learning (DRL). Next, the same method is applied for routing energy at three different layers, including smart home, Home Area Network (HAN), and grid layers. Finally, an attack detection framework is proposed to detect False Data Injection (FDI) attacks and electricity theft using Long Short-Term Memory (LSTM) and Convolutional Neural Network (CNN)  \cite{h10,e2}.

\section{Problem Statement and Motivation}
The origination of IoE emanates from the high penetration of RESs across the power grids and can be used for energy management enhancement. IoE provides a decentralized structure that facilitates accommodation and integration of RESs besides increasing demand-supply reliability. Real-time monitoring that is enabled through bidirectional communication is another significant advantage of this cloud-based network. Finally, automation controls implemented in an IoE-based power network improve energy efficiency, leading to profit maximization  \cite{k8,h11}.
 The residential sector, including both consumers and prosumers, is known as the most important sector in the IoE networks and economic goals are one of the main motivations for this sector  \cite{h13}. IoE facilitates handling challenges associated with OS optimization to improve energy efficiency, profit maximization, and decreasing GHGs emissions. Besides that, IoE can be used to enhance energy routing optimization. As the core of IoE, an Energy Router (ER) adjusts the energy route dynamically among the devices considering technical and economic constraints to minimize energy loss and maximize profit  \cite{e6, 12}. Consequently, optimizing electricity routing is vital, either made for a single user or at the Home Area Network (HAN) level. 
Despite the above-mentioned advantages, IoEs are vulnerable to cyber-attacks due to the broad range of bidirectional interconnection among installed smart devices. Therefore, to take advantage of this technology, assuring security against malicious cyber/physical attacks is required  \cite{h12}.
Several investigations have been conducted on the application of IoE during the last decade, but the challenges mentioned above have not been addressed well.  \cite{14,e3} proposed a real‐time power scheduling method based on the structural design of the IoE. The suggested approach has a high computational burden and highly relies on regional zones. Also, there is no specific planning for the home appliances operation, which is necessary to discover surplus electricity for market trading and profit maximization. In [16], a scheduling method for smart homes and plug-in hybrid electric vehicles (PHEVs) is introduced. Unfortunately, the RESs are ignored in this study, and the utilized Support Vector Machine (SVM) algorithm can assign a binary function to the devices.
Furthermore, the centralized decision-making structure that is employed may result in decreasing efficiency and arduous recovery after disasters or attacks. A two-stage device scheduling with dynamic programming is proposed by  \cite{15}. In this study, only major renewable energy stations are considered, with no scheduling at the home level. None of the previous works included comprehensive scheduling considering both energy units and home appliances to find surplus energy. Almost all of the previous works consider both consumers and prosumers with the same role in the market as price takers, resulting in a non-transparent market since a large share of the network belongs to the household users. An intelligent scheduling program would make users prone to act as a price maker in the electricity market, leading to a competitive marketplace. Dealing with the uncertainties of RESs, handling sudden changes in the consuming or generating patterns, and computation complexity also need to be addressed.
After developing a scheduling algorithm that satisfies all technical and economic goals at the residential level, the next step is optimizing the energy route among all devices in a smart home and the HANs. In  \cite{h16} a hierarchical optimization method for ERs is proposed to reduce the complexity of centralized optimal dispatch on large-scale systems. The consideration level is higher than the sub-distribution network, where there is no routing algorithm for the residential section. Also, storage and Electric Vehicles (EVs) effects have been ignored. An energy router design with an optimized algorithm that covers smart homes, HANs, and grid levels at the same time is still absent. Moreover, the desired algorithm that can perform in online and offline mode has not been studied yet, and islanding mode identification should also be investigated.
IoE relies on two-way communication with a fully interconnected Advanced Metering Infrastructures (AMIs) network, which makes the system vulnerable to cyber and physical malicious activities.  For instance, a cyber-attack on the Supervisory Control and Data Acquisition (SCADA) system of a regional electricity distribution company in Ukraine led to a power outage for 225,000 customers, which caused massive technical and economic damages  \cite{20,e5}. Also, physical tampering and bypassing are responsible for 20\% of the electricity lost in India  \cite{19,e4}. Therefore, guaranteeing security comes with high priority in IoE-enabled power systems.
In  \cite{h18}, a defence mechanism based on an interval state predictor is proposed, which can mitigate the effect of malicious attacks. However, the proposed method does not consider the storage, EVs and DC busses in the ERs. The security challenges of integrating fog computing into the IoE are studied in [22], focusing on collision attacks. Still, FDI or physical attacks like AMIs manipulation and bypassing are not considered. Therefore, developing a comprehensive algorithm for cyber and physical attack detection in different layers of IoE is necessary.

\section{	Research Objectives}
Three main objectives are specified for this research proposal, including optimizing OS, designing an optimized ER, and improving the cyber-physical security of the IoE. The OS algorithm makes adaption with IoE-based network profitable for both prosumers (by maximizing profit) and utilities (by minimizing cost). Moreover, consumers are prone to reduce the bill amount via flatting the load curve. Energy routing is the next step for energy management utilizing IoE, which must be done in both users and grid-level in offline and online modes. Lastly, cyber-physical security is needed to assure the system's security via FDI and physical attacks detection. The main objectives of this research proposal are as follows: 
\subsection{	Optimizing operation scheduling of energy components}
The leading goal of this research is to address the challenges originated from the aggregation of RESs via IoE for efficient energy management while reducing the cost and GHGs emissions and establishing a competitive market. By developing and maintaining a scheduling algorithm, on the one hand, all consumers make a profit. These earnings come from incentives and tariff adjusting (for shifting consumption to off-peak hours), besides selling the surplus energy to the market. On the other hand, utilities can reduce the operation, transmission, and maintenance costs, reducing the need for new investigations. Ultimately, this research's first objective is to develop a scheduling algorithm to increase the customer's profit by managing their consumption and trading identified surplus electricity.
\subsection{	Energy routing optimization}
Once the operation times for each energy unit in the system have been scheduled, the subsequent critical step is optimizing the energy routing. This process involves using an energy routing algorithm to plan the amount and direction of electricity flow, taking into account the type of power, whether alternating current (AC) or direct current (DC). This planning must adhere to both technical and economic constraints to ensure efficient and cost-effective energy distribution.

The primary goal of the energy routing algorithm is to optimize the distribution of electricity across the network. This involves determining the most efficient pathways for energy to travel from its generation points to its various consumption points. The algorithm must consider several factors, such as the capacity of transmission lines, the demand at different nodes, and the cost associated with transmitting energy over various distances. Additionally, it must address the differences in handling AC and DC power, which have distinct characteristics and requirements.
Once the initial routing plan is created, it is sent back to the scheduling algorithm. Here, alternative routing and scheduling options are evaluated based on customer preference levels. This iterative process ensures that the final plan not only meets technical and economic criteria but also aligns with customer needs and priorities. The aim is to find the best trade-off between scheduling and routing to maximize overall system profit while maintaining reliability and efficiency.
The development of a sophisticated energy routing algorithm is the second key objective of this research proposal. This algorithm will leverage advanced mathematical models and computational techniques to handle the complexity of modern energy systems. By integrating real-time data and predictive analytics, the algorithm can dynamically adjust to changes in demand and supply, thereby optimizing performance continuously.
Energy routing optimization is a pivotal component of managing modern energy systems. It requires a nuanced approach that balances technical constraints, economic considerations, and customer preferences. The proposed research will focus on creating an advanced energy routing algorithm that enhances the efficiency and profitability of energy distribution. By doing so, it will contribute significantly to the development of smarter, more resilient energy networks capable of meeting the evolving demands of today's energy landscape. This research aims to deliver innovative solutions that can be applied in real-world scenarios, providing a substantial impact on the efficiency and sustainability of energy systems.
\subsection{	Cyber-Physical attack detection}
As previously highlighted, Internet of Energy (IoE) networks are susceptible to cyber-attacks, including false data injection (FDI) attacks, as well as physical tampering. These threats can severely impact the performance of scheduling and routing algorithms, which rely heavily on the accuracy of the data they receive. Consequently, ensuring the security of these networks is crucial. The third objective of this research proposal is to conduct a comprehensive security analysis aimed at safeguarding data integrity and protecting against cyber-physical attacks. This objective will involve developing robust mechanisms to detect and mitigate potential threats, thereby enhancing the resilience and reliability of IoE systems \cite{19}.
\subsection{	Research Contributions}
The ultimate goals of this research are to develop a comprehensive framework that includes optimized scheduling and routing algorithms along with an advanced attack detection mechanism to identify FDI attacks and electricity theft. To achieve these objectives, two Deep Reinforcement Learning (DRL)-based algorithms will be developed. These algorithms will address the first two objectives by mitigating the uncertainties associated with Renewable Energy Sources (RES) and reducing the complexity and computational burden of the problem.
Given the critical importance of data accuracy in ensuring the reliable performance of scheduling and routing algorithms, the research proposes the use of Convolutional Neural Networks (CNN) and Long Short-Term Memory (LSTM) networks for detecting FDI attacks. These advanced machine learning techniques are well-suited for analyzing complex data patterns and identifying anomalies that may indicate a cyber-attack. Furthermore, a Deep Convolutional Neural Network (DCNN) model is proposed for detecting electricity theft. The DCNN model will leverage its deep learning capabilities to accurately identify patterns associated with unauthorized electricity usage, thereby preventing financial losses and ensuring the integrity of the energy distribution system.
The contributions of this research are multifaceted and significant. By developing optimized scheduling and routing algorithms, the research aims to enhance the efficiency and reliability of energy systems. The introduction of sophisticated attack detection mechanisms, using cutting-edge machine learning techniques, will ensure the security and integrity of the data, safeguarding IoE networks from both cyber and physical threats. This comprehensive approach will not only improve the performance of energy systems but also contribute to the development of more resilient and secure IoE infrastructures, capable of meeting the evolving challenges of the modern energy landscape. The research contribution can be summarized as follows:
\section{	Developing a DRL-based scheduling framework for techno-economic improvement in the power network}
As the first major contribution, this research proposal introduces a Q-Learning based framework, an off-policy Reinforcement Learning (RL) algorithm, to optimize operation schedules within diverse environments and under various constraints. These environments include home appliances, energy resources (such as generation and storage units), and the technical, economic, and communication layers that interconnect them. The proposed framework aims to create an optimal scheduling program that transforms traditional consumers into prosumers by shifting the operation of flexible loads to off-peak hours. This shift not only reduces energy costs for households but also fosters a competitive market where even those without their own energy generation units can participate effectively.
The scheduling of Renewable Energy Sources (RESs), fixed storage systems, and Electric Vehicles (EVs) is contingent on their specific limitations and the overall conditions of the interconnected environments. These conditions include the dynamic variations in energy supply and demand, as well as the technical and economic constraints that can fluctuate over time. By leveraging Deep Reinforcement Learning (DRL), the proposed framework enables consumers to act as prosumers. This is achieved by strategically shifting flexible loads to off-peak periods, thereby lowering energy bills and allowing for the trading of surplus energy.
In practical terms, the Q-Learning based framework optimizes the scheduling process by continuously learning and adapting to changes in the environment. It considers the real-time status and capabilities of home appliances, energy generation and storage units, and the overall energy market conditions. By doing so, it identifies the most cost-effective and efficient times to operate various devices and systems, ensuring that energy consumption is aligned with periods of lower demand and higher availability of renewable energy.
Moreover, the framework's adaptability to varying technical and economic constraints ensures that it remains effective even as external conditions change. This adaptability is crucial in a dynamic energy landscape where supply and demand can be highly variable. The use of DRL techniques allows the system to handle these complexities and make informed decisions that maximize both individual and collective benefits.
The Q-Learning based optimization framework represents a significant advancement in energy management. It not only enhances the efficiency of energy use but also empowers consumers to become active participants in the energy market. By optimizing the scheduling of energy resources and shifting loads to off-peak hours, the framework reduces costs, increases the utilization of renewable energy, and supports the development of a more resilient and sustainable energy system. This research aims to provide a robust solution that addresses the current challenges in energy management and contributes to the evolution of smart, integrated energy networks.
\subsection{	Developing an energy routing algorithm }
via DRL considering technical and economic constraints
Routing optimization extremely relies on OS since the amount and direction of energy that needs to be sent or received among devices is planned based on the scheduling algorithm's results. Also, the transmission constraints that are notified by utilities must be satisfied. As the second contribution, a DRL based algorithm is proposed to deal with the complexity of the problem and minimize the negative impact of power generation and consumption uncertainty at both user-level and grid levels.  Furthermore, the proposed DRL-based OS algorithm is capable of handling islanding mode challenges besides offline working ability.
\subsection{	Cyber-physical attack detection guarantying information correctness}
To ensure the security of the IoE, an LSTM algorithm is proposed to detect FDI attacks in time series data, combined with a CNN for feature extraction and correlation identification among different types of data to deal with uncertainties.
Additionally, an Ensemble Deep Convolutional Neural Network (EDCNN) algorithm is proposed for Electricity Theft Detection (ETD) caused by AMIs' physical manipulation. As the first layer of the model, a random under bagging technique is applied to deal with the imbalance data, and then deep CNNs are utilized on each subset. Finally, a voting system is embedded in the last part.
\subsection{	Significance of the Study}
Electrical energy plays a crucial role in this era and as a high demand type of energy. Utilizing RESs is vital due to global warming caused by GHGs emissions originated from using fossil fuels to generate electricity. High penetration of RESs requires modern energy management systems to deal with technical and economic constraints. IoE is the enabler for modern energy management, which provides novel energy trading features besides real-time monitoring and dynamic scheduling. 
After reviewing all the related studies, this proposal determined the research gaps of application of IoE for energy management, including optimal OS, optimal energy routing, security, and energy trading. Consequently, a comprehensive scheduling program is proposed to optimize the energy components' operation scheduling considering techno-economic constraints besides achieving the best energy routing plan. Also, a reliable platform is required to ensure the correctness of the information used in the scheduling and routing algorithms. Therefore, an attack detection algorithm is developed detecting FDI attacks and electricity theft  \cite{k7,k8}.

\section{Propsoed	Methodology} 
After reviewing the various studies and reports, along with industrial and market demand, the problem statements have been recognized. A comprehensive literature review then demonstrated the research gaps and challenges that have led to defining research objectives and contributions. Next, the best solution has been suggested for every goal where several datasets have been collected to train, evaluate, and test the proposed algorithms. The proposed solutions are as follows:
\subsection{	Deep Reinforcement Learning based scheduling algorithm}
The Internet of Energy (IoE) is a concept that enhances energy management by integrating renewable energy sources and distributed generators (DGs). It facilitates a bidirectional flow of information and real-time monitoring, leading to significant improvements in energy efficiency. The operation scheduling of energy units in smart homes is a critical component of IoE because the residential sector accounts for a substantial portion of global energy consumption. Effective scheduling can optimize energy use, reduce costs, and enhance the overall efficiency of energy systems. Reinforcement Learning (RL) approaches are particularly well-suited for this task due to their ability to generalize and resist bias. These algorithms can handle incomplete information by extracting useful data from the available inputs. Moreover, RL algorithms typically have a fast convergence rate to approximate solutions, which reduces their complexity and computational burdens compared to other machine learning algorithms. This efficiency makes RL an attractive choice for managing the dynamic and often uncertain environments associated with energy systems. As illustrated in Figure 6, an RL algorithm comprises two main components: an agent and an environment that interact frequently. The agent selects actions from an action space and sends signals to the environment, learning how to behave in each state. Each time an action is taken, the environment generates a reward value. This reward, along with the new state, is sent back to the agent for the next iteration. The reward indicates whether an action is positive or negative, while the new state provides the agent with the necessary information for future decisions.
RL algorithms excel in handling complex problems with uncertainties because they can iteratively correct errors during the training process. This adaptability is crucial in energy management, where conditions can change rapidly and unpredictably. By continuously learning and adjusting, RL algorithms can optimize energy scheduling, ensuring that energy consumption aligns with periods of low demand and high availability of renewable resources.
In the context of smart homes, RL-based scheduling can transform traditional energy consumers into prosumers—entities that both consume and produce energy. By shifting flexible load operations to off-peak hours, households can reduce energy costs and participate in energy trading markets, even without their own energy generation units. This shift not only benefits individual households by lowering bills but also contributes to a more balanced and efficient energy grid.
The IoE concept, supported by RL approaches, offers a powerful framework for improving energy efficiency through optimized scheduling and real-time management. By leveraging the strengths of RL algorithms, this research aims to enhance the operational efficiency of smart homes and integrate them more effectively into the broader energy system. This approach promises to deliver significant economic and environmental benefits, paving the way for a more sustainable and resilient energy future.

\begin{figure}[!h]
 
\includegraphics[width=.5\textwidth]{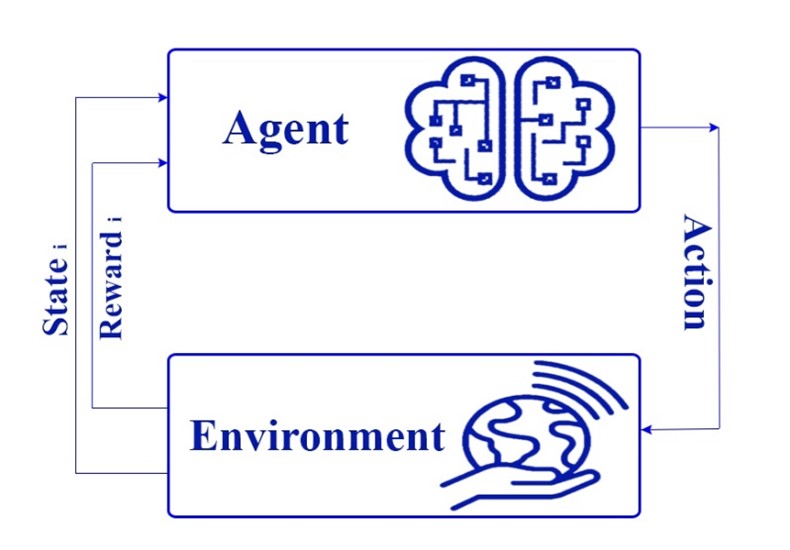}

 \caption{ \centering Reinforcement Learning algorithm Design}
 \label{Figure:Results_anomalies}
\end{figure}

Aside from their advantages, RL algorithms suffer from dimensionality problems similar to other ML algorithms \cite{h2}. Besides that, assume a Markovian property for the environment, which is not the case all the time.   To address these drawbacks, DRL is introduced. Utilizing DRL improves the control strategy and optimization process via feature extraction and alleviate the dimensionality problem by ignoring redundant information due to the uncertainties.
This research presents a DRL based framework to optimize scheduling in the IoE-network under different environment and limitations. As Figure 7 demonstrates, the environments include the application layer (home appliances, energy resources), technical layer (e.g. stability, reliability, power quality, ampacity, etc.), economic layer (e.g. generation and transmission costs, energy price, tariff adjustment, etc.), and communication layers (e.g. Automated meter reading, wired and wireless networks). An optimal scheduling program can transform customers into prosumers by shifting the flexible load operation time to off-peak hours, which leads to a competitive market. Scheduling of resources, fixed storages, and EVs depend on their intrinsic limitations and rely on the overall condition of other environments. Also, technical, and economic constraints may vary, considering dynamic changes in energy supply and demand.
\begin{figure}[!h]
 
\includegraphics[width=.5\textwidth]{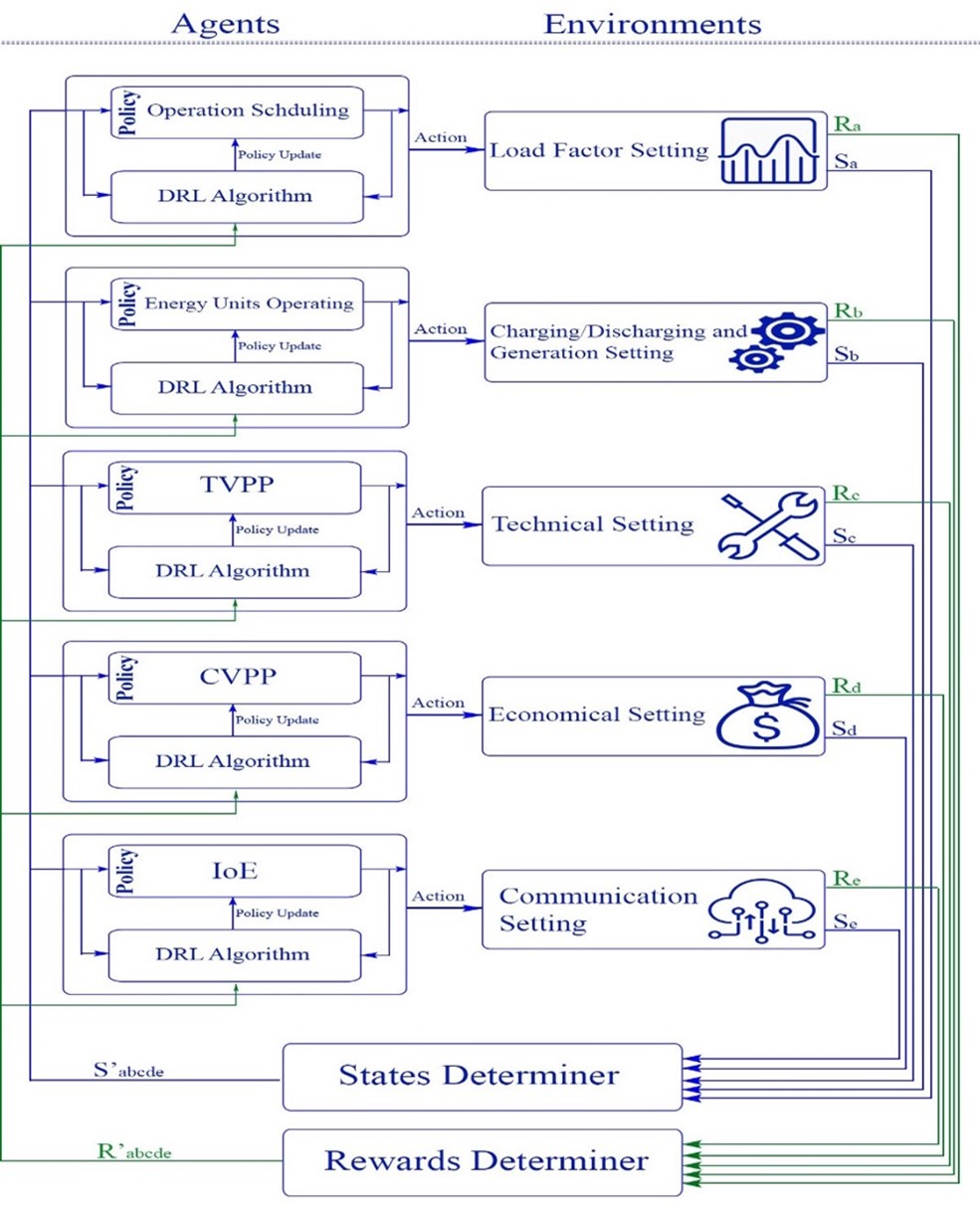}

 \caption{ \centering The structure of the Proposed Method}
 \label{Figure:3}
\end{figure}

\subsection{Q-learning based routing algorithm}
As an off-policy RL algorithm, Q-learning intends to discover the best action to take based on the current state while learning the best value function maximizing the total reward. In the first step, the algorithm observes the environment and decides how to act based on the current state. Then, a reward or penalty is assigned, while the algorithm is trained based on the last experiment. Finally, after several iterations, the algorithm reaches an optimal strategy. 
As figure 8 shows, three different routing levels include nano-grid, microgrid, and grid levels.

\begin{figure}[!h]
\includegraphics[width=.5\textwidth]{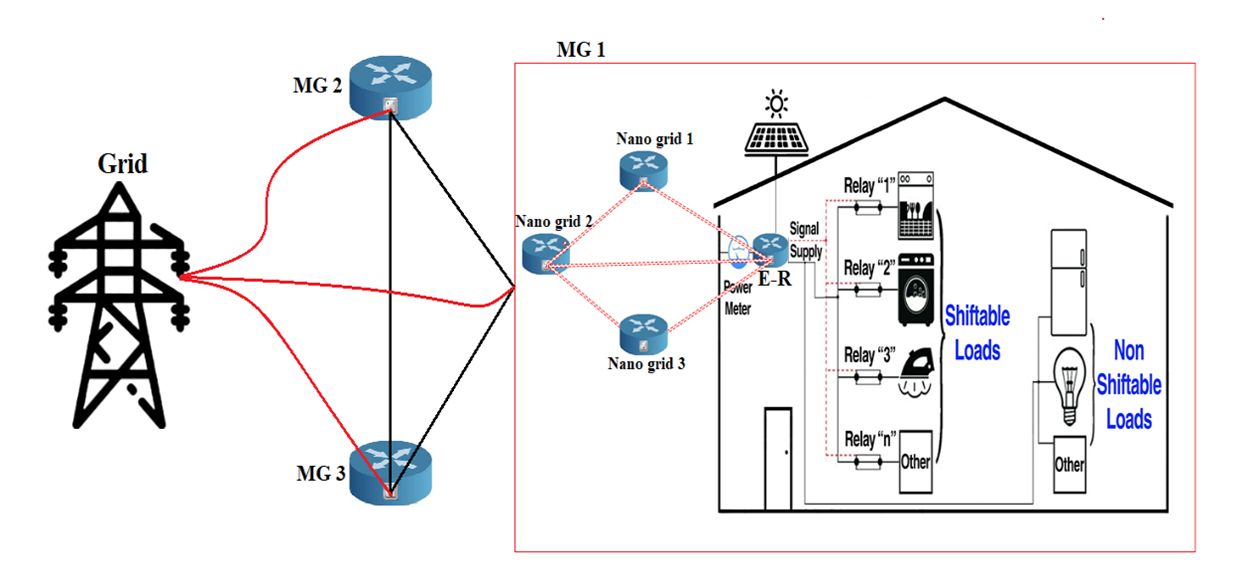}
 \caption{ \centering Schematic of different levels in the routing algorithm}
 \label{Figure2}
\end{figure}

Considering the number of energy units and different levels of considerations, the Q-learning well situs method for electricity routing in the IoE-based power systems since the algorithm must route electricity among energy units at the home level and at HAN and grid levels.
\subsection{	Joint LSTM and CNN for attack detection}
The proposed attack detection in this research has two levels, including cyber and physical malicious activities discovery. For the cyber layer, considering the sequential nature of the data, an LSTM model is proposed to detect FDI attacks in the network to ensure the correctness of the data used in the scheduling and routing algorithms. 
Also, to detect electricity theft, as one of the major physical attacks, CNN is proposed. The information used for ETD can be considered a multidimensional dataset since various interpretation takes place based on the different points of view. As a deep learning algorithm, CNN can extract high-level features capturing the spatial and temporal dependencies. 
As figure 9 shows, the proposed framework has three main stages: 1) data preprocessing to classify datasets and extract meaningful features, 2) training the model based on the datasets, and 3) detecting the malicious activity as FDI attack to electricity theft.

\begin{figure*}[!h]
\includegraphics[width=.5\textwidth]{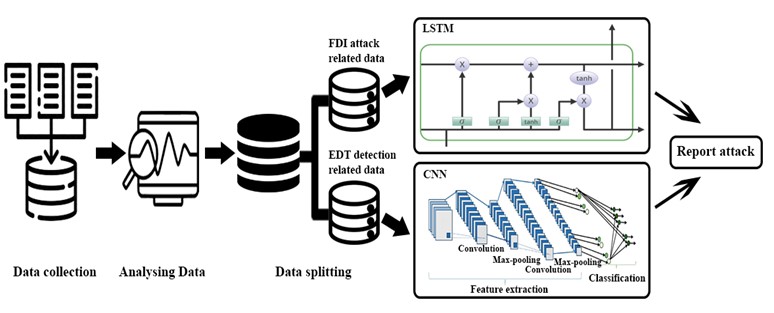}

 \caption{ \centering Attack detection Mechanism}
 \label{Figurel}
\end{figure*}

\section{	Research Contributions}
 This research's ultimate goals are to propose a framework, including optimum scheduling and routing algorithms, besides an attack detection mechanism to discover FDI attacks and electricity theft. To achieve the objectives as mentioned earlier, two DRL based algorithms will be developed for the first two objectives to mitigate the effect of RES's uncertainties and reduce the complexity of the problem and computational burden. Considering the importance of data accuracy, CNN and LSTM algorithms are proposed for FDI attack detection, and Deep Convolutional Neural Network (DCNN) model is proposed for electricity theft detection. The research contribution can be summarized as follows:
\subsection{	Developing a DRL-based scheduling framework for techno-economic improvement in the power network}
The first major contribution of this research proposal is the introduction of a Q-Learning based framework, an off-policy Reinforcement Learning (RL) algorithm, designed to optimize operation schedules within various environments and constraints. These environments encompass home appliances, energy resources (including generation and storage units), and the technical, economic, and communication layers that interconnect them.
The proposed framework aims to create an optimal scheduling program that transforms traditional energy consumers into prosumers by shifting the operation of flexible loads to off-peak hours. This shift reduces energy costs for households and fosters a competitive market where even those without their own energy generation units can participate effectively. By participating in this market, households can trade surplus energy, contributing to the overall efficiency and sustainability of the energy system.
The scheduling of Renewable Energy Sources (RESs), fixed storage systems, and Electric Vehicles (EVs) within this framework depends on their specific limitations and the overall conditions of the interconnected environments. These conditions include dynamic variations in energy supply and demand, as well as the technical and economic constraints that can fluctuate over time. By leveraging Deep Reinforcement Learning (DRL), the proposed framework enables consumers to act as prosumers. This is achieved by strategically shifting flexible loads to off-peak periods, thereby lowering energy bills and allowing for the trading of surplus energy.
In practical terms, the Q-Learning based framework optimizes the scheduling process by continuously learning and adapting to changes in the environment. It considers the real-time status and capabilities of home appliances, energy generation and storage units, and the overall energy market conditions. By doing so, it identifies the most cost-effective and efficient times to operate various devices and systems, ensuring that energy consumption is aligned with periods of lower demand and higher availability of renewable energy.
Moreover, the framework's adaptability to varying technical and economic constraints ensures that it remains effective even as external conditions change. This adaptability is crucial in a dynamic energy landscape where supply and demand can be highly variable. The use of DRL techniques allows the system to handle these complexities and make informed decisions that maximize both individual and collective benefits.
The Q-Learning based optimization framework represents a significant advancement in energy management. It not only enhances the efficiency of energy use but also empowers consumers to become active participants in the energy market. By optimizing the scheduling of energy resources and shifting loads to off-peak hours, the framework reduces costs, increases the utilization of renewable energy, and supports the development of a more resilient and sustainable energy system. This research aims to provide a robust solution that addresses the current challenges in energy management and contributes to the evolution of smart, integrated energy networks.

 \cite{e6}.
\subsection{	Developing an energy routing algorithm }via DRL considering technical and economic constraints
Routing optimization extremely relies on OS since the amount and direction of energy that needs to be sent or received among devices is planned based on the scheduling algorithm's results. Also, the transmission constraints that are notified by utilities must be satisfied. As the second contribution, a DRL based algorithm is proposed to deal with the complexity of the problem and minimize the negative impact of power generation and consumption uncertainty at both user-level and grid levels.  Furthermore, the proposed DRL-based OS algorithm is capable of handling islanding mode challenges besides offline working ability.
\subsection{	Cyber-physical attack detection guarantying information correctness}
To ensure the security of the IoE, an LSTM algorithm is proposed to detect FDI attacks in time series data, combined with a CNN for feature extraction and correlation identification among different types of data to deal with uncertainties.
Additionally, an Ensemble Deep Convolutional Neural Network (EDCNN) algorithm is proposed for Electricity Theft Detection (ETD) caused by AMIs' physical manipulation. As the first layer of the model, a random under bagging technique is applied to deal with the imbalance data, and then deep CNNs are utilized on each subset. Finally, a voting system is embedded in the last part.

\section{Conclusion}
In conclusion, this research addresses the critical challenges of energy management in IoE-enabled smart grids by developing advanced algorithms for operation scheduling, energy routing, and cyber-physical security. The DRL-based scheduling algorithm optimizes the use of RESs, reducing costs and emissions while maximizing economic benefits for prosumers. The energy routing algorithm ensures efficient electricity flow, adapting to technical and economic constraints. Furthermore, the proposed security framework enhances the resilience of IoE networks against cyber and physical attacks, ensuring reliable operation. The contributions of this research provide a robust foundation for future developments in smart grid technology, promoting the integration of renewable energy and enhancing overall energy management efficiency. These advancements are essential for meeting the growing energy demands in an environmentally sustainable manner. By leveraging cutting-edge AI technologies, this research paves the way for smarter, more resilient power systems. The findings highlight the potential for IoE to revolutionize energy management, making it a cornerstone for future innovations in the field.

\bibliographystyle{IEEEtran}
\bibliography{ref}

\end{document}